\documentclass[twocolumn,showpacs,superscriptaddress]{revtex4}
\usepackage{amssymb}
\usepackage{amsmath}
\usepackage{graphicx}
\usepackage{epsfig}
\usepackage{amsfonts}

\begin{document}

\title{ Non-Adiabatic Fluctuation in Measured Geometric Phase }
\author{Qing Ai}
\affiliation{Department of Physics, Tsinghua University, Beijing 100084, China}
\author{Wenyi Huo}
\affiliation{Department of Physics, Tsinghua University, Beijing 100084, China}
\author{Gui Lu Long}
\affiliation{Department of Physics, Tsinghua University, Beijing 100084, China}
\affiliation{Tsinghua National Laboratory for Information Science and Technology, Beijing
100084, China}
\author{C. P. Sun}
\affiliation{Institute of Theoretical Physics, Chinese Academy of Sciences, Beijing,
100080, China}

\begin{abstract}
We study how the non-adiabatic effect causes the observable fluctuation in
the ``geometric phase" for a two-level system, which is defined as the
experimentally measurable quantity in the adiabatic limit. From the Rabi's
exact solution to this model, we give a reasonable explanation to the
experimental discovery of phase fluctuation in the superconducting circuit
system [P. J. Leek, \textit{et al}., Science \textbf{318}, 1889 (2007)],
which seemed to be regarded as the conventional experimental error.
\end{abstract}

\pacs{ 03.65.Vf, 03.65.Ca, 03.65.Ta }
\maketitle

\textit{Introduction.}- It was discovered by I. I. Rabi
\cite{Rabi37} that the non-adiabatic transition of a quantum system
in a time-dependent magnetic field was subject to the sign of its
magnetic momentum. The exact solution was first given in 1937, but
its physical significance for the relative phase acquired under
adiabatic evolution was not clarified until five decades later
\cite{Berry84}. It was M. V. Berry who found that this phase might
contain a geometric part, now called the Berry's phase. Then the
quantum adiabatic approximation theorem (QAAT) \cite{Messiah62} was
reproved to naturally include the Berry's phase \cite{Sun88a} and
generalized to deal with the non-adiabatic effects for many cases
\cite{Sun88b,Sun90a,Sun90b,Sun93,Sun95}. On the other hand, because
of its geometric dependence, conditional geometric phase was
proposed as an
intrinsically fault-tolerant way of performing quantum computation \cite%
{Jonathan00}.

In this paper, associated with a recent experiment about phase fluctuation
in the superconducting circuit system \cite{Leek07}, the above Rabi's
solution is used to study in details the non-adiabatic effects for a
two-level system (TLS) in a harmonically rotated field (see Fig.1). This
field can be realized with a microwave field perpendicular to the static
magnetic field both applied to the system. With the phase of the microwave
linearly varying with time, the Hamiltonian harmonically rotates in the
parametric space.

Generally speaking, the Berry's phase is always accompanied with the
dynamical phase, and thus its pure effect can not be observed directly.
However, we apply a $\pi$-pulse to the TLS so that the evolution is divided
into two parts with both of them in the same path but in the opposite
directions. In this case, the effect of the dynamical phase can be
completely eliminated. This is the technique referred to spin echo technique
\cite{Abragam61}. Thus, the pure geometric effect can be observed. And that
may result in observable fluctuation of the measured geometric phase due to
the Rabi's non-adiabatic transitions.

\begin{figure}[ptb]
\includegraphics[bb=20 450 560 830,width=8.5 cm]{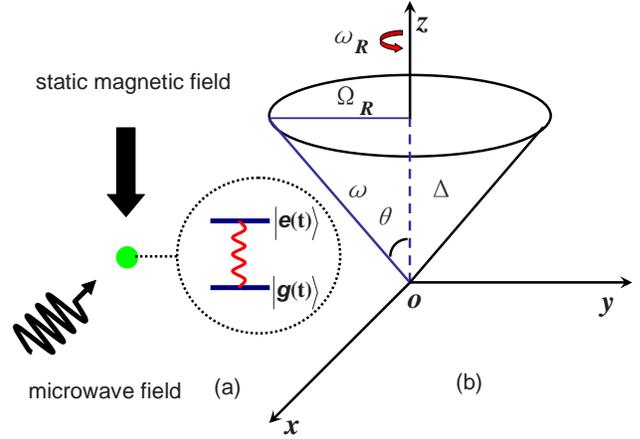}
\caption{ (color online) (a) Schematic diagram of a TLS in a static magnetic
field along with a microwave field. (b) Realized effective Hamiltonian
rotating in the parametric space. }
\label{Sketch}
\end{figure}

\textit{Non-Adiabatic Effect with Berry's Phase.}-The evolution of the
system can be well described with the Hamiltonian
\begin{equation}
H(t)=\frac{1}{2}(\Delta\sigma_z+\Omega_R\sigma_x\cos\omega_Rt+
\Omega_R\sigma_y\sin\omega_Rt) \text{,}  \label{H}
\end{equation}
where $\Delta$ is the energy splitting without the microwave field, $\Omega_R
$ the Rabi frequency of the microwave, $\omega_R$ the oscillating frequency
of the microwave's phase, $\sigma_{x,y,z}$ the Pauli matrixes. Note that
Hamiltonian (\ref{H}) is exactly the effective Hamiltonian realized in Ref.%
\cite{Leek07}, where the rotating wave approximation \cite{Scully} was
applied. Straightforwardly, its instantaneous eigen states are obtained as
\begin{align}
|e(t)\rangle&=\cos\frac{\theta}{2}|0\rangle+\sin\frac{\theta}{2}%
e^{i\omega_Rt}|1\rangle \text{,} \\
|g(t)\rangle&=\sin\frac{\theta}{2}e^{-i\omega_Rt}|0\rangle-\cos\frac{\theta}{%
2}|1\rangle \text{,}  \label{EigenState}
\end{align}
with corresponding eigen energies $\pm\omega/2$. Here the energy splitting
is $\omega=\sqrt{\Delta^2+\Omega_R^2}$, and the mixing angle is $%
\theta=\tan^{-1}(\Omega_R/\Delta)$. We also emphasize that due to the
requirement of the single valueness of the eigenfunctions for a given
Hamiltonian without singularity, the phase factor $\exp(\pm i\omega_Rt)$ in $%
|e(t)\rangle$ or $|g(t)\rangle$ is fixed once the factor in the
other state is chosen \cite{Shapere89}.

At time $t$, the evolution state is assumed to be a superposition
$|\psi (t)\rangle =\alpha (t)|e(t)\rangle +\beta (t)|g(t)\rangle $
of  two
instantaneous eigen states . The time-dependent Schr\"{o}dinger equation $%
H|\Psi (t)\rangle =i\partial _{t}|\Psi (t)\rangle $ leads to the following
equations of coefficients
\begin{align}
\dot{\alpha}& =-i(\frac{\omega }{2}+\omega _{R}\sin ^{2}\frac{\theta }{2}%
)\alpha +i\beta ^{\prime }\frac{\omega _{R}}{2}\sin \theta \text{,}
\label{alpha} \\
\dot{\beta ^{\prime }}& =i(\frac{\omega }{2}-\omega _{R}\cos ^{2}\frac{%
\theta }{2})\beta ^{\prime }+i\alpha \frac{\omega _{R}}{2}\sin \theta \text{,%
}  \label{beta}
\end{align}%
where $\beta ^{\prime }(t)=\beta (t)\exp (-i\omega _{R}t)$.

Under the adiabatic conditions
\begin{align}
\frac{\omega }{2}+\omega _{R}\sin ^{2}(\frac{\theta }{2})& \gg \frac{\omega
_{R}}{2}\sin \theta \text{,}  \label{AdiabaticCondition1} \\
\frac{\omega }{2}-\omega _{R}\cos ^{2}(\frac{\theta }{2})& \gg \frac{\omega
_{R}}{2}\sin \theta \text{,}  \label{AdiabaticCondition2}
\end{align}%
the adiabatic approximate solutions to Eqs.(\ref{alpha}) and
(\ref{beta}) is obtained \ by ignoring the terms with $\omega
_{R}\sin \theta $ $/2$ . They show that both norms of the amplitudes
remain the same as their initial values, while they acquire a
Berry's geometric phase $\pm \omega _{R}t(1-\cos \theta)/2$ in
addition to the dynamical phase $\pm \omega t/2$ respectively.

On the other hand, the above Eqs.(\ref{alpha}) and (\ref{beta}) can be
solved exactly and it should be done so when the adiabatic condition is
broken under certain circumstances. Thus, we have
\begin{align}
\alpha & =A_{1}e^{i\omega _{+}t}+A_{2}e^{i\omega _{-}t}\text{,}
\label{AlphaExact} \\
\beta ^{\prime }& =B_{1}e^{i\omega _{+}t}+B_{2}e^{i\omega _{-}t}\text{,}
\label{BetaExact}
\end{align}%
where $\omega _{\pm }=(-\omega _{R}\pm \Omega )/2\equiv (-\omega _{R}\pm
\sqrt{\omega ^{2}-2\omega \omega _{R}\cos \theta +\omega _{R}^{2}})/2$, and
the coefficients are determined by the initial values $\alpha (0)$ and $%
\beta (0)$,
\begin{align}
A_{1}& =\frac{\beta (0)\omega _{R}\sin \theta +\alpha (0)(-\omega
+\Sigma_+
)}{2\Omega }\text{,}  \notag \\
A_{2}& =\frac{-\beta (0)\omega _{R}\sin \theta +\alpha (0)(\omega +\Sigma_- )}{%
2\Omega }\text{,}  \notag \\
B_{1}& =\frac{\beta (0)(\omega +\Sigma_- )+\alpha (0)\omega _{R}\sin \theta }{%
2\Omega }\text{,}  \notag \\
B_{2}& =\frac{\beta (0)(-\omega +\Sigma_+ )-\alpha (0)\omega _{R}\sin \theta }{%
2\Omega }\text{,}  \notag
\end{align}%
where $\Sigma_\pm =\omega _{R}(1\pm\cos \theta )+2\omega _{+}.$

\begin{figure}[ptb]
\includegraphics[bb=95 267 479 573,width=8.5 cm]{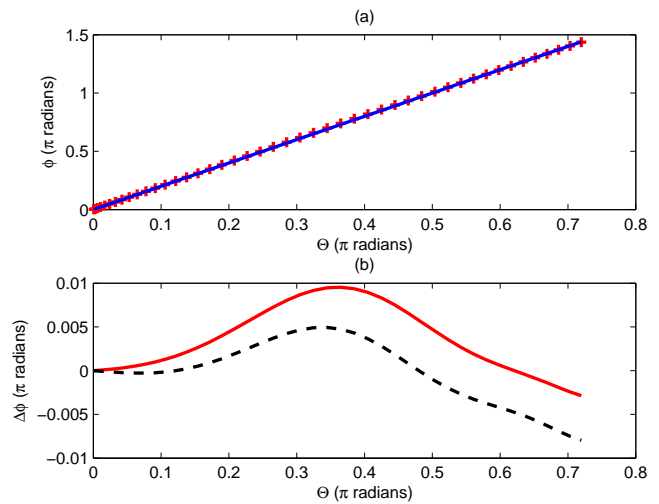}
\caption{ (color online) (a) Comparison between the Berry's phase (blue
solid line) $\protect\phi_B$ and non-adiabatic phase (red cross) $\protect%
\phi_{na}$ with $n=1$ circular rotation in each round. (b) The discrepancy
between them with red solid line for the numerical result and black dashed
line for the second order approximation. }
\label{ExactAngle}
\end{figure}

Generally speaking, the Berry's phase can not be observed directly from the
experiment, since the dynamical phase always occurs along with the Berry's
phase. Here, we consider a concrete case where the total evolution is
divided into two rounds, both of which are in the same path but with
opposite directions. Provided that both the excited and the ground states
acquire the same dynamical phase but with opposite signs in each round, the
dynamical phase can be canceled when the amplitudes are exchanged after the
first round of evolution. Thereafter, by experiencing the inverse rotation
in the parametric space, the dynamical phase is canceled while the Berry's
phase is doubled since the latter depends on the sign of the angle velocity $%
\omega_R$ while the former does not. Excellent agreement may be expected
with the Berry's prediction
\begin{equation}
\phi_B=2\omega_RT(1-\cos\theta)  \label{phiB}
\end{equation}
provided that the adiabatic condition, i.e.,
Eq.(\ref{AdiabaticCondition1}), is satisfied. Here, $T$ is the
evolution time for each round.

However, when the non-adiabatic effect is considered, small deviation is
expected. In Fig.(\ref{ExactAngle}), we plot the exact phase calculated from
Eqs.(\ref{AlphaExact}) and (\ref{BetaExact}), which is defined as  the phase
of $\alpha (2T)\beta ^{\ast }(2T)$, denoted as
\begin{equation}
\phi _{na}=\text{angle}(\alpha (2T)\beta ^{\ast }(2T))\text{.}
\end{equation}%
This is an observable quantity in experiment, which can be
determined by measuring the complex amplitudes  $\alpha (2T)$ and $\
\beta (2T).$ It just recovers the Berry's phase in adiabatic limit.

It is predicted that the Berry's phase is proportional to the solid angle $%
\Theta =2\pi (1-\cos \theta )$ subtended by the path. Recently, a
measurement of the Berry's phase in the superconducting qubit was carried
out \cite{Leek07}. In order to compare the theoretical analysis with the
experimental result, we adopt the same parameters as those given in Ref.\cite%
{Leek07}, i.e., $\Delta =50$ MHz, $\omega _{R}=(4n+1)$ MHz with $n$ being
the number of loops. The tiny difference between those two can almost not be
distinguished in Fig.(\ref{ExactAngle}a). Moreover, as shown in Fig.(\ref%
{ExactAngle}b), there are small oscillations in the deviation between them, $%
\Delta \phi =\phi _{na}-\phi _{B}$, with the root-mean-square deviation of $%
0.015$ rad from the expected lines while the counterpart for $n=1.5$ is $%
0.043$ rad. They are in reasonable agreement with the experimental result,
i.e., $0.14$ rad \cite{Leek07}, considering that the rotating wave
approximation \cite{Scully} was applied to obtain the Hamiltonian (\ref{H})
and the non-adiabatic effect in the process of applying the microwave field
also accounted for part of the deviation.

\textit{Second-Order Fluctuation.}-It was C. N. Yang who firstly pointed out
that the Berry's phase could be recovered from the original QAAT by
retaining the first order term $O(\omega _{R}/\omega )$ in the phase. And
this point of view was shortly confirmed by one (CPS) of the authors \cite%
{Sun95} using the Rabi's exact solution. Here, with careful calculation to
the second order term $O(\omega _{R}/\omega )^{2}$, the fluctuation in the
phase is obtained. For an initial state with $\alpha (0)=\beta (0)=1/\sqrt{2}
$, the measured deviation from the expected Berry's phase is given as
\begin{align}
\Delta \phi =& \lambda ^{2}(\sin \phi _{1}+4\sin \phi _{2}+2\sin \phi
_{3}+2\sin \phi _{4}  \notag \\
& +2\sin \phi _{5}+\sin \phi _{6}+2\sin \phi _{7}+\sin \phi _{8}  \notag \\
& +2\sin \phi _{9}+2\sin \phi _{10}+4\sin \phi _{11}+4\sin \phi _{12})\text{,%
}  \label{DeltaPhi}
\end{align}%
where $\lambda =\omega _{R}\sin \theta /2\omega $, $\phi _{j}=\phi
_{j}^{\prime }-\phi _{B}$ ($j=1,2\ldots $) with
\begin{align}
\phi _{1}^{\prime }& =\pi -2T\omega _{R}\cos \theta \text{,}  \notag \\
\phi _{2}^{\prime }& =-T\omega _{R}\text{,}  \notag \\
\phi _{3}^{\prime }& =T\omega _{R}(-1+2\cos \theta )\text{,}  \notag \\
\phi _{4}^{\prime }& =-\frac{T(\omega _{R}^{2}+8\omega _{R}\omega +4\omega
^{2}-4\omega _{R}\omega \cos \theta -\omega _{R}^{2}\cos 2\theta )}{4\omega }%
\text{,}  \notag \\
\phi _{5}^{\prime }& =\pi +\frac{T(-(\omega _{R}+2\omega )^{2}+4\omega
_{R}\omega \cos \theta +\omega _{R}^{2}\cos 2\theta )}{4\omega }\text{,}
\notag \\
\phi _{6}^{\prime }& =\pi +\frac{T((\omega _{R}-2\omega )^{2}-\omega
_{R}^{2}\cos 2\theta )}{2\omega }\text{,}  \notag \\
\phi _{7}^{\prime }& =\frac{T(\omega _{R}^{2}-2\omega _{R}\omega +4\omega
^{2}-\omega _{R}^{2}\cos 2\theta )}{2\omega }\text{,}  \notag \\
\phi _{8}^{\prime }& =\pi +\frac{T(\omega _{R}^{2}+4\omega ^{2}-\omega
_{R}^{2}\cos 2\theta )}{2\omega }\text{,}  \notag \\
\phi _{9}^{\prime }& =\pi +\frac{T((\omega _{R}-2\omega )^{2}-4\omega
_{R}\omega \cos \theta -\omega _{R}^{2}\cos 2\theta )}{4\omega }\text{,}
\notag \\
\phi _{10}^{\prime }& =\frac{T(\omega _{R}^{2}+4\omega ^{2}-4\omega
_{R}\omega \cos \theta -\omega _{R}^{2}\cos 2\theta )}{4\omega }\text{,}
\notag \\
\phi _{11}^{\prime }& =\frac{T(\omega _{R}^{2}-8\omega _{R}\omega +4\omega
^{2}+4\omega _{R}\omega \cos \theta -\omega _{R}^{2}\cos 2\theta )}{4\omega }%
\text{,}  \notag \\
\phi _{12}^{\prime }& =\pi +\frac{T((\omega _{R}-2\omega )^{2}+4\omega
_{R}\omega \cos \theta -\omega _{R}^{2}\cos 2\theta )}{4\omega }\text{.}
\notag
\end{align}

\begin{figure}[ptb]
\includegraphics[bb=95 267 479 573,width=8.5 cm]{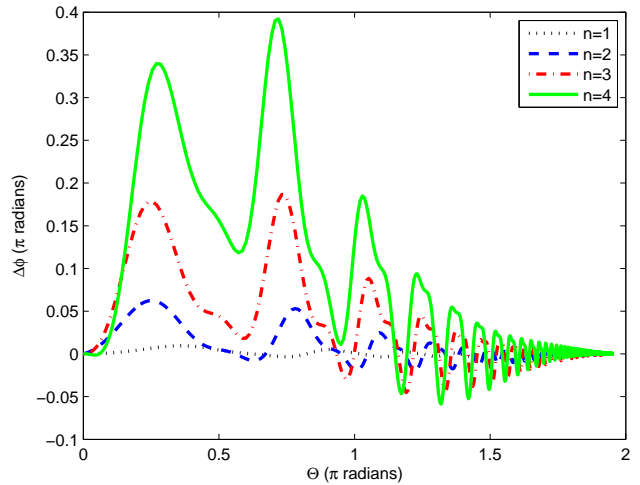}
\caption{ (color online) Comparison between different numbers of rotation
loops $n$, with black dotted line for $n=1$, blue dashed line for $n=2$, red
dash-dot line for $n=3$ and green solid line for $n=4$. }
\label{ComparisonNs}
\end{figure}

It can be seen from Eq.(\ref{DeltaPhi}) that the sinusoidal deviation from
the expectation will be present in the measured phase. As shown in Fig.\ref%
{ExactAngle}(b), the behavior is well described by the second order
approximation. Although there is small distinction between it and the exact
result, it is believed that it is due to the higher order terms. Based on
the above theoretical analysis, we may safely arrive at the conclusion that
due to the non-adiabatic evolution of the TLS, not only do the relative
phases of the amplitudes of the two states change with time, but also do
their norms. In other words, the non-adiabatic transition results in the
small fluctuation of the measured phase.

To further explore the non-adiabatic effect, we investigate the phase
fluctuation for different rotation velocity $\omega_R$'s. In Fig.\ref%
{ComparisonNs}, the discrepancy between the measured phase and the Berry's
phase is plotted. As expected from Eq.(\ref{DeltaPhi}), the sinusoidal
oscillation is again witnessed. Additionally, the oscillating amplitudes
rise as the rotating velocity is increased. Here, the total time for
evolution is fixed \cite{AboutTime}. This result is consistent with Eq.(\ref%
{DeltaPhi}) as $\Delta\phi$ scales as $\omega_R^2$. Notice that all of them
nonexceptionally approach zero at the both ends. It is a reasonable result
since $\Delta\phi$ vanishes as $\theta=0$. Actually, the Hamiltonian remains
the same as its initial state in the parametric space. The only effect for
time evolution is to acquire a dynamical phase. On the other hand, as $\theta
$ approaches $\pi/2$,
\begin{equation}
\omega=\sqrt{\Omega_R^2+\Delta^2}>\Omega_R\gg\Delta>\omega_R  \notag
\end{equation}
since $\Delta$ is fixed. Due to vanishing of $\lambda$, we have a zero
deviation from the predicted phase, i.e., $\Delta\phi=0$. In the limit $%
\theta\rightarrow\pi/2$, the initial state $|\psi(0)\rangle=(|0\rangle+|1%
\rangle)/\sqrt{2}$ is the eigen state of the initial Hamiltonian $%
H(0)\simeq\Omega_R\sigma_x/2$. Intuitionally, the evolution of the system is
similar to the situation that a classical magnetic moment initially parallel
to the applied field closely follows the rotation of the field.

\textit{Conclusion.}-In this paper, we have investigated the fluctuation in
the phase due to the non-adiabatic evolution. In contrast to the adiabatic
evolution, the sinusoidal deviation from the expected line drawn for the
Berry's phase is observed. The phase fluctuations discovered in the recent
experiment can be  partially explained by our theoretical analysis. For a
given time of evolution, it is predicted that the fluctuation from the
Berry's prediction rises larger and larger as the number of loops is
increased.

\textit{Acknowledgement.}-One (QA) of the authors would be grateful
for stimulating discussions with Y. S. Li and valuable comments on
the manuscript from Z. Q. Yin. This work is partially supported by
the National Fundamental Research Program Grant No. 2006CB921106,
China National Natural Science Foundation Grant No. 10775076.


\smallskip

\begin{thebibliography}{99}
\bibitem{Rabi37} I. I. Rabi, Phys. Rev. \textbf{51}, 652 (1937).

\bibitem{Berry84} M. V. Berry, Proc. R. Soc. London Ser. A \textbf{392}, 45
(1984).

\bibitem{Messiah62} A. Messiah, Quantum Mechanics (North-Holland, Amsterdam,
1962).

\bibitem{Sun88a} C. P. Sun, J. Phys. A \textbf{21}, 1585 (1988).

\bibitem{Sun88b} C. P. Sun, Phys. Rev. D \textbf{38}, 2908 (1988).

\bibitem{Sun90a} C. P. Sun, Phys. Rev. D \textbf{41}, 1318 (1990).

\bibitem{Sun90b} C. P. Sun and M. L. Ge, Phys. Rev. D \textbf{41}, 1349
(1990).

\bibitem{Sun93} C. P. Sun, Phys. Scr. \textbf{48}, 393 (1993).

\bibitem{Sun95} C. P. Sun and L. Z. Zhang, Phys. Scr. \textbf{51}, 16 (1995).

\bibitem{Berry87} M. V. Berry, Proc. R. Soc. London Ser. A \textbf{414}, 31
(1987).

\bibitem{Jonathan00} J. A. Jones, V. Vedral, A. Ekert, and  G. Castagnoli,
Nature \textbf{403}, 869 (2000).

\bibitem{Shapere89} A. Shapere, F. Wilczek, Geometric Phases in Physics
(World Scientific, Singapore, 1989).

\bibitem{Abragam61} A. Abragam, Principles of Nuclear Magnetism (Oxford
Univ. Press, Oxford, 1961).

\bibitem{Leek07} P. J. Leek, J. M. Fink, A. Blais, R. Bianchetti, M. G\"{o}%
ppl, J. M. Gambetta, D. I. Schuster, L. Frunzio, R. J. Schoelkopf, A.
Wallraf, Science \textbf{318}, 1889 (2007).

\bibitem{Scully} M. O. Scully, M. S. Zubairy, \textit{Quantum Optics},
(Cambridge University Press, Cambridge, England, 1997).

\bibitem{AboutTime} In the experiment, the total time consisted of the
evolution of two rotations and the processes of applying and removing the
microwave field. For the sake of excluding the phase decoherence, the total
time for evolution $T^\prime=500$ ns is fixed. In contrast, the time for
each rotation $T=nT^\prime/(2n+1/2)$ varies with $n$.
\end{thebibliography}
\end{document}